\documentclass[sn-mathphys,Numbered]{sn-jnl}% Math and Physical Sciences Reference Style
\usepackage{graphicx}%
\usepackage{multirow}%
\usepackage{amsmath,amssymb,amsfonts}%
\usepackage{mathrsfs}%
\usepackage[title]{appendix}%
\usepackage{xcolor}%
\usepackage{textcomp}%
\usepackage{manyfoot}%
\usepackage{booktabs}%
\usepackage{algorithm}%
\usepackage{algorithmicx}%
\usepackage{algpseudocode}%
\usepackage{listings}%
\usepackage{array}
\newcolumntype{L}[1]{>{\raggedright\let\newline\\\arraybackslash\hspace{0pt}}p{#1}}
\newcolumntype{C}[1]{>{\centering\let\newline\\\arraybackslash\hspace{0pt}}p{#1}}
\newcolumntype{R}[1]{>{\raggedleft\let\newline\\\arraybackslash\hspace{0pt}}p{#1}}

\usepackage[bottom]{footmisc}

\usepackage{amssymb}
\usepackage{hyperref}
\usepackage{multirow}
\usepackage{multicol}
\usepackage{float}
\usepackage{amsthm}
\usepackage{amsmath}
\usepackage{rotating} 
\usepackage{graphicx}
\usepackage{makecell}
\usepackage{tabularx}
\usepackage{xcolor,graphicx,float}  
\usepackage{multirow}

\begin{document}

\title{How do transportation professionals perceive the impacts of AI applications in transportation? A latent class cluster analysis}

%%=============================================================%%
%% Prefix	-> \pfx{Dr}
%% GivenName	-> \fnm{Joergen W.}
%% Particle	-> \spfx{van der} -> surname prefix
%% FamilyName	-> \sur{Ploeg}
%% Suffix	-> \sfx{IV}
%% NatureName	-> \tanm{Poet Laureate} -> Title after name
%% Degrees	-> \dgr{MSc, PhD}
%% \author*[1,2]{\pfx{Dr} \fnm{Joergen W.} \spfx{van der} \sur{Ploeg} \sfx{IV} \tanm{Poet Laureate} 
%%                 \dgr{MSc, PhD}}\email{iauthor@gmail.com}
%%=============================================================%%

\author[1]{\fnm{Yiheng} \sur{Qian}}\email{yihengqian@ufl.edu}

\author[2]{\fnm{Tejaswi} \sur{Polimetla}}\email{tpolimetla@ufl.edu}

\author[3]{\fnm{Thomas} \sur{W. Sanchez}}\email{tom.sanchez@vt.edu}

\author*[1]{\fnm{Xiang} \sur{Yan}}\email{xiangyan@ufl.edu}

\affil[1]{\orgdiv{Department of Civil and Coastal Engineering}, \orgname{University of Florida}, \orgaddress{\city{Gainesville}, \postcode{32603}, \state{Florida}, \country{United States}}}

\affil[2]{\orgdiv{Department of Geography}, \orgname{University of Florida}, \orgaddress{\city{Gainesville}, \postcode{32603}, \state{Florida}, \country{United States}}}

\affil[3]{\orgdiv{Department of Urban Affairs and Planning}, \orgname{Virginia Tech}, \orgaddress{\city{Arlington}, \postcode{22203}, \state{Virginia}, \country{United States}}}

%%==================================%%
%% sample for unstructured abstract %%
%%==================================%%

\abstract{%Over half of the respondents currently have a neutral stance on AI applications in transportation, and respondents with positive views slightly outnumber those with negative views. Factor analysis is employed to derive three latent factors: AI's efficiency benefits, AI's equity concerns, and AI's equity benefits. 
Recent years have witnessed an increasing number of artificial intelligence (AI) applications in transportation. As a new and emerging technology, AI's potential to advance transportation goals and the full extent of its impacts on the transportation sector is not yet well understood. As the transportation community explores these topics, it is critical to understand how transportation professionals, the driving force behind AI Transportation applications, perceive AI's potential efficiency and equity impacts. Toward this goal, we surveyed transportation professionals in the United States and collected a total of 354 responses. Based on the survey responses, we conducted both descriptive analysis and latent class cluster analysis (LCCA). The former provides an overview of prevalent attitudes among transportation professionals, while the latter allows the identification of distinct segments based on their latent attitudes toward AI. We find widespread optimism regarding AI's potential to improve many aspects of transportation (e.g., efficiency, cost reduction, and traveler experience); however, responses are mixed regarding AI's potential to advance equity. Moreover, many respondents are concerned that AI ethics are not well understood in the transportation community and that AI use in transportation could exaggerate existing inequalities. Through LCCA, we have identified four latent segments: \textit{AI Neutral}, \textit{AI Optimist}, \textit{AI Pessimist}, and \textit{AI Skeptic}. The latent class membership is significantly associated with respondents' age, education level, and AI knowledge level. Overall, the study results shed light on the extent to which the transportation community as a whole is ready to leverage AI systems to transform current practices and inform targeted education to improve the understanding of AI among transportation professionals.  }


%%\pacs[JEL Classification]{D8, H51}

%%\pacs[MSC Classification]{35A01, 65L10, 65L12, 65L20, 65L70}

\maketitle

\section{Introduction}

Artificial intelligence (AI) is becoming an integral part of our daily lives, bringing changes to a variety of industries such as healthcare, marketing, finance, and transportation. In a July 2020 report, the U.S. Department of Transportation Intelligent Transportation System (ITS) Joint Program Office identified 60 AI-enabled applications in ITS across 11 categories, such as advanced driver assistance systems, traveler decision support tools, and asset management, covering various aspects of transportation that affect the lives of almost all travelers \citep{vasudevan2020identifying}. Moreover, AI technologies are believed to be capable of addressing ITS operational changes and transportation needs in a wide range of real-life scenarios, such as urban arterial networks, multimodal corridors, and underserved communities \citep{vasudevan2020real}.

Since AI is still a new and emerging technology, the potential of AI systems to advance transportation goals and the full extent of AI’s impact on the transportation sector are largely unknown. While AI systems hold great potential to improve transportation for communities and travelers, potential bias in AI development and deployment may exacerbate existing transportation inequalities. As the transportation community navigates the path forward, it can expect to engage in ongoing debates and discussions concerning the benefits of AI, the challenges of its implementation, and the ethical implications that arise. These discussions are already happening in many domains (e.g., facial recognition) and the broader scientific community. For example, a recent survey conducted by Nature on researchers ($N > 1,600$)  indicated that scientists are both excited and concerned by the increasing use of AI tools in research \citep{van2023ai}. More than half of the respondents believed that AI has enhanced data processing and computation speeds, thereby saving researchers time and money. Yet, at the same time, over half of them expressed concerns that AI results might reinforce biases or discriminatory data. The mixed feelings expressed by the surveyed researchers imply that AI can be a double-edged sword for transportation: While deploying AI in transportation can lead to significant benefits, these applications may cause ethical concerns and be accompanied by unintended consequences such as widening inequalities. 

%\textbf{Particularly noteworthy is the realm of transportation where the influence of AI on efficiency and equity remains uncertain. This ambiguity is crucial to address, especially considering that transportation professionals directly} drive traffic policies, planning, and operations. Understanding their perspectives on AI's efficiency and equity has become paramount. Consequently, gaining in-depth insights into the opinions of professionals in the transportation sector is essential for a comprehensive assessment of both the potential and limitations of AI in this field.

%\textbf{In some scenarios where AI has been preliminarily applied, people have widely observed improvements in efficiency.} However, simultaneously, there has been a \textbf{growing chorus of complaints} regarding the equity of AI. 

While there has been a growing interest among transportation professionals to learn about different aspects of AI, we have little knowledge of how the community as a whole perceives the potential use of AI in transportation as well as its impact. The existing AI applications in transportation are mostly driven by technology developers and early adopters who are more receptive to innovation and eager to explore the potential of new technologies. Meanwhile, as recent surveys have consistently shown that about half or more of Americans are skeptical of autonomous vehicles \citep[e.g.,][]{gross2022consumer}, a major domain of AI application more familiar to the transportation community, we can project that some individuals have doubts about the use of AI in transportation. However, there is a lack of empirical work to shed light on whether and to what extent a wide range of opinions toward AI exist among transportation professionals, as well as the reasons behind the contrasting views. In-depth research on these disparities is needed for the transportation community to gain a better understanding of the challenges that might be encountered when promoting AI technology in the transportation sector. Moreover, there is a lack of research to assess transportation professionals’ AI knowledge level, their perceptions of AI's impacts on the transportation system, and their willingness and capacity to leverage AI systems to transform current transportation practices. Such work is needed because how the transportation community as a whole perceives AI and its efficiency and equity impacts will significantly affect whether and how fast these technologies are adopted by transportation agencies. Also, the current level of awareness and knowledge of AI technologies and AI applications in transportation would determine the transportation workforce's readiness to manage AI systems being deployed in the real world. 
%Additionally, this research can help identify effective methods for AI widespread adoption.

Motivated by these research needs, we mainly address two research questions in this study: How do transportation professionals perceive the efficiency and equity impacts of AI-enabled applications in transportation? What distinctive population segments exist concerning these perceptions? Toward this purpose, we surveyed transportation professionals in North America (mostly in the U.S.) to understand their perception of the potential impacts of AI's applications in transportation. The respondents were also asked to evaluate some attitudinal statements about equity and ethical considerations for AI applications in transportation. In addition, it asks about respondents’ knowledge of and training in AI and their sociodemographic information.

To our knowledge, this is the first survey on this topic and it reflects the early stages of significant innovation adoption within the transportation field. This paper provides survey results from transportation professionals concerning the future of AI, and latent class cluster analysis is used to segment respondents into respondent groups having distinctive views of AI's efficiency and equity impacts. The insights gained from understanding the sociodemographic characteristics and attitudes of each latent segment are instrumental in crafting personalized communication strategies, as well as guiding policy and strategy formulation. These insights are especially valuable in addressing any resistance or uncertainty towards AI within certain sub-groups. Moreover, the paper sheds light on potential shifts in attitudes toward AI in the transportation sector as demographic dynamics evolve.

\section{Literature Review}
\subsection{AI applications in transportation}

The 21st century has seen the rapid development of new technologies across industries, with artificial intelligence being one of the hallmarks. The field of transportation is no exception, and there is vast potential for the deployment of AI in a plethora of capacities. Some key applications in transportation include traveler decision support tools, transportation systems management and operations (TSMO), transit operations and management, and asset management. Traveler decision support tools use AI information about a transportation network \citep{walker_artificial_2020} to help travelers plan trips that fit their needs and preferences. Machine learning comprises much of the research in this area \citep{jiang_cellular_2022}, and applications include improvements in travel time predictions by Google Maps \citep{derrow-pinion_eta_2021} and use in airports for predicting congestion, analyzing air traffic control speech, and detecting irregularities in flight paths and taxiing \citep{tien_roles_2022}. TSMO refers to the maintenance and improvement of transportation infrastructure and operations rather than capacity expansion, while transit operations and management is a subset of TSMO which focuses on transit systems. Applications include locating and tracking incidents, adaptive ramp metering to manage congestion, signal coordination, timetable optimization \citep{muller-hannemann_estimating_2022}, route optimization \citep{ge_artificial_2021}, and fare enforcement. Finally, asset management refers to the upkeep of physical infrastructure to ensure good and safe operating conditions while balancing costs. Existing examples include automating rail inspection and pavement condition detection \citep{tsai_successful_2023}. 

With these applications come many potential benefits to using AI in transportation systems. On the organizational level, benefits could include improved efficiency, decreased costs, and better environmental outcomes. Dynamic scheduling algorithms \citep{lv_ai_2021}, real-time route optimization \citep{iyer_ai_2021}, and value chain transformation via the physical internet \citep{nikitas_artificial_2020} could all lead to improvements in efficiency. Demand forecasting and responsiveness \citep{abduljabbar_applications_2019} and infrastructure monitoring \citep{okrepilov_modern_2022} have the potential to reduce operational costs. Decreased congestion \citep{hasan_review_2019} and vehicle reduction \citep{rigole__2014} can improve environmental outcomes. Travelers can expect better safety, accessibility, and convenience. Crime forecasting \citep{kouziokas_application_2017} and adaptability to road and traffic conditions \citep{boukerche_artificial_2020} could make transit a safer experience. Computer vision-enabled cooperative traffic signal assistance \citep{yang_cooperative_2022}, LiDAR-enabled infrastructure assessment \citep{ai_automated_2016}, and haptic feedback technologies \citep{boldini_inconspicuous_2021} have the potential to expand accessibility. Finally, convenience can be enhanced by minimizing wait times \citep{yin_review_2020}, optimizing routes \citep{adler_toward_1998}, and creating smart recommendations and navigation for tourists \citep{tsaih_artificial_2018}. 
 
\subsection{AI's equity impacts}

However, these benefits are not without potential equity concerns. The ultimate objective of transportation equity is to provide equal access to social and economic opportunity by providing equitable access across communities and population groups \citep{sanchez_moving_2003}. They address a wide range of socioeconomic and geospatial inequities in transportation. Historically, transportation policies in the United States have favored highway development over public transit, leading to a plethora of negative consequences. For one, highways were often constructed through minority communities as a part of “slum clearance” and “urban renewal” in the 1950s and 1960s. These projects disrupted community life, and continue to contribute to increased pollution and impaired health in marginalized communities. Highway development also encourages housing development further away from city centers, exacerbating residential segregation and income inequalities. This led to a “spatial mismatch” of jobs as jobs on the outskirts of cities became inaccessible to those living in city centers. The disparity in transportation investments between highways and public transit and uneven urban development patterns have left many without options except for driving, making transportation costs constitute greater proportions of household expenditures for low-income households than for higher-income households. These are but some of the many inequities stemming from historic investments and the continual shaping of transportation policies. 

The introduction of AI to the transportation system could either work to remedy these inequities or exacerbate them further. As discussed in \cite{vasudevan2020identifying}, there many potential AI applications that can enhance the provision of accessible, equitable, reliable, and affordable transportation services for traditionally underserved travelers. Examples include AI-enhanced citizen engagement, AI-enabled routing and wayfinding tools for pedestrians, AI-enabled payment assistance, and AI-powered assistive robots for people with disabilities. Nevertheless, rather than addressing the needs of underserved communities and population groups, the existing AI applications in transportation are mostly geared toward enhancing driver assistance systems, mitigating traffic congestion, and automating infrastructure assessments \citep{vasudevan2020real, iyer_ai_2021}. Moreover, much of the emphasis for these AI applications is on evaluating the potential of AI systems to improve or replace current transportation practices, with little attention paid to the ethical and equity implications of AI deployment. However, as demonstrated in other domains such as facial recognition \citep{najibi2020racial}, potential bias in the AI development and deployment processes risks certain population groups such as racial minorities. For example, AI-based decision-support systems can lead to policies and decisions that leave out the needs of people with disabilities if they are underrepresented in the data used to train the AI. In other words, if ethical and equity considerations are not carefully accounted for, AI technologies designed to improve transportation processes and outcomes could often end up with the unintended consequences of exacerbating existing inequalities.

%designed to improve efficiency could have unintended consequences in terms of negative equity impacts if these concerns are not taken into serious consideration. 

%AI systems have been increasingly deployed in transportation to mitigate traffic congestion. Examples of applications include scanning traffic patterns to optimize traffic signal controls, detecting roads with urgent maintenance needs to reduce road accidents (the main contributor to traffic congestion), and recommending personalized and convenient shared mobility options for travelers to reduce the use of single-occupancy vehicles.} So far, limited research has focused on the equity implications of deploying AI technologies to transform transportation practices. 

\subsection{Transportation professionals’ perception of AI}
At present, studies on how professionals working in the transportation sector perceive AI's potential for transportation applications and its impacts are limited. The most relevant body of work is research on how the general public perceives autonomous vehicles (AVs), a crucial research field of AI and one of the most familiar AI applications in transportation. \citet{ othman2021public} reviewed studies on public acceptance and perception of AVs: older adults are pessimistic about AVs, despite expectations of increased accessibility for this group; also, males and highly educated individuals express have more positive attitudes towards AVs compared to females and those with lower education levels. A survey study conducted in Taiwan \citep{chen2020preparing} delves into societal preparedness for AVs by surveying AI experts and people majoring in computer science or electrical engineering. Although AI experts highlighted AVs' positive impact on disadvantaged communities' mobility, both groups viewed social equity issues as less important and urgent than cybersecurity and data privacy. A survey of Brisbane residents \citep{butler2021factors} revealed significant variations in attitudes regarding the benefits of AVs. Positive attitudes were observed among young and middle-aged adults, people with disabilities, and public transport users, while Gender showed no significant correlation with attitudes toward AVs. 

Some studies further segmented the population into different groups based on their perceptions or attitudes. A study conducted by \citet{hilgarter2020public} in Carinthia (Austria) in September 2018 classified respondents into four groups based on their attitudes towards autonomous vehicles: rejectors (10.5\%), conservatives (26.3\%), pragmatists (26.3\%), and enthusiasts (36.8\%). This classification was determined by the researchers based on statements provided during interviews, and the main limitation is that the sample size was very small (N=19). The latent class cluster analysis (LCCA), a quantitative approach, has been used in past studies to investigate differences in how people perceive automated vehicles (AVs). For instance, one study looked at the perceived advantages and disadvantages of AVs and classified users into the following groups: AV(-inclined) over walk/bike, AV over flight, zero-occupant AV over occupied AV, and AV over transit \citep{kim_identifying_2019}. Another study examined perceptions of the autonomous taxi market and formed the following categories: neutral and diverse travelers, conservative and strict travelers, and open and enjoying travelers \citep{dai_future_2023}. LCCA is also widely used as a statistical method in other transportation topics: \citet{ton2020latent} revealed different daily mobility travel patterns, \citet{wang2022latent} studied the heterogeneity among riders using on-demand service to connect to light rail stations, \citet{lee2020millennials} explored gradual changes of multimodality across age and generation. By grouping respondents using LCCA, the sociodemographic and behavioral characteristics of each group can be further examined and studied. 

%Given the many unknowns that come with any new methodologies, it is key that practitioners in the field of transportation \textbf{have a clear understanding of a path forward. }There has been limited research conducted about transportation professionals’ views on AI adoption in transportation, \textbf{and the potential equity implications that could arise.} Depending on their stance on AI, professionals may take on various roles - a pragmatist, an antagonist, or an enthusiast \cite{neri_role_2020}. Additionally, when they \textbf{perpetuate messages of risk}, it is generally with regards to \textbf{counterfactuals} rather than real-life incidents \cite{neri_role_2020}. Despite these early findings, existing work in the field is limited to a narrow selection of early adopters, and there is no clear understanding of the \textbf{workforce development that would be needed and the equity concerns that could arise.} 

%This study employs latent class cluster analysis (LCCA) to understand the diverse perspectives and views of the transportation community. 

Existing AI applications in transportation are driven by a small group of early adopters, widespread AI adoption requires engagement from the entire transportation community. Therefore, it is essential to understand how the transportation community perceives AI as the field navigates a path forward in a time with many unknowns with this new technology. Without a clear grasp of how professionals perceive the development of AI adoption in transportation, it is difficult to appropriately plan for changes in the labor force, take full advantage of potential benefits for society at large, or safeguard against potential ethical and equity challenges. To address these research gaps, we conducted a survey to understand how professionals in the field of transportation perceive the potential of AI use in transportation, its benefits, and the equity challenges that may arise with AI adoption. Using the data, a latent class analysis was conducted to categorize transportation professionals based on their attitudes and perceptions.

\section{Data and Approach}
\subsection{Survey}
We designed a survey to investigate how transportation professionals in the United States perceive the potential of AI and its potential impacts. The survey questionnaire (attached in the Appendix) was divided into four sections: respondents’ perception of AI’s impact on transportation, their knowledge level and training in AI, their perception of AI’s equity and ethical concerns, and their sociodemographic and economic information. The survey consists of 23 questions, including 18 close-ended multiple-choice questions, 3 matrix table questions, and two open-ended questions (one for respondents to type in their thoughts on equity/ethics of using AI in transportation or the other for general comments). Since AI and transportation are both broad concepts that can be defined broadly, the following definitions are provided in the survey: \\

\begin{itemize}
\item Artificial Intelligence (AI) refers to processes that make it possible for systems to replace or augment routine human tasks or enable new capabilities that humans cannot perform. AI enables systems to (1) sense and perceive the environment, (2) reason and analyze information, (3) learn from experience and adapt to new situations, potentially without human interaction, and (4) make decisions, communicate, and take action.

\item Transportation mainly refers to transportation planning and engineering practices that facilitate the movement of people and goods.\\
\end{itemize}

A pilot survey was first conducted among a small group of individuals (mostly transportation survey experts), whose feedback was incorporated into the final survey. Our survey targeted a wide range of transportation professionals working in both public and private sectors, regardless of whether they actively engage in conducting AI-related work or not. Specifically, to engage professionals working in the public sector, the research team has gathered a list of email addresses consisting of employees from state departments of transportation (DOTs), county and city DOTs, metropolitan planning organizations (MPOs), and transit agencies from 48 contiguous U.S. states. For each agency, we collected the email addresses of two to three individuals who serve in leadership positions related to research and innovation, planning/engineering, and civil rights. We requested these individuals to fill out the survey and forward it to their colleagues. We have also advertised the survey on the TMIP listserv, an online Community of Practice for the travel and freight modeling and planning communities. Moreover, we reached out to the Institute of Transportation Engineers and its local chapters (e.g., Florida Puerto Rico District ITE), several Transportation Research Board Standing Committees (e.g., AED50), and the transportation planning division of the American Planning Association; some of these organizations promote the survey through their email lists or on their newsletters. In addition, to maximize participation from minority transportation professionals, we have emailed the leaders of the local chapters of the Conference of Minority Transportation Officials and requested them to help adverse the service. Finally, we have engaged some professionals working in consulting through personal networks. 

The survey collection efforts occurred between January 2023 to May 2023. No incentives were provided for survey participation. In the end, we collected a total of 359 responses. Out of these, 270 responses met the requirements for modeling (i.e., almost complete), with 253 being complete responses. The medium response time is 9 minutes. We are not able to compute the response rate for the survey because of the non-probability sampling approach (i.e., convenience sampling) used here. We did not apply probability sampling here mainly because there is no clear definition of who counts as a transportation professional, which means that there is not a defined ``population" for this study and so we would not be able to measure whether the survey sample is representative or not. Accordingly, our primary goal in the participant recruitment process was to reach a more \textit{diverse} set of transportation professionals; to achieve this purpose, as discussed above, we reached out to a wide range of transportation agencies and professional organizations across the continental United States.

\subsection{Method}

%Our goal is to identify the characteristics of each preference segment and the differences in sociodemographic, AI knowledge level, and education among these segments. 

To address the two research questions discussed above, we employ a combination of descriptive analysis and statistical modeling. Regarding descriptive analysis, we summarize the results from relevant survey questions and use cross-tabulation to explore the relationship between respondent's perceptions of AI and their personal characteristics. We then use the latent class cluster analysis (LCCA) to further segment respondents into distinctive groups based on their perception of AI's efficiency and equity impacts. 

%explore people's underlying preferences, we complement descriptive analysis with modeling. Compared to descriptive analysis, modeling can provide more in-depth and concise results.

LCCA, a probabilistic-based clustering method, is commonly used to identify population segments with similar preferences and latent attitudes. The LCCA approach comprises two sub-models: the \textbf{membership model} and the \textbf{measurement model}. The LCCA model can be expressed as:

\begin{equation}
f_2(\boldsymbol{y}_n \mid \boldsymbol{z}_n) = \sum_{k=1}^K P(k \mid \boldsymbol{z}_n) f_1(\boldsymbol{y}_n \mid k)
\end{equation}

where \(n\) is the case subscript, \(k\) is a nominal latent variable, \(Zn\) is a vector of covariates, \(yn\) is a vector of indicators, \(P\left(k \mid \boldsymbol{z}_n\right)\) is the membership probability for a certain latent class given covariates, \(f_1\left(\boldsymbol{y}_n \mid k\right)\) is the probability density of \(yn\) given \(k\), and \(f_2\left(\boldsymbol{y}_n \mid \boldsymbol{z}_n\right)\) is the probability density of \(yn\) given \(zn\).

Applied to the sample as a whole, a useful interpretation of Eq. (1) is that the model aims to delineate the set of latent classes (as represented by the \(P\left(k \mid \boldsymbol{z}_n\right)\) that will best explain the joint distribution of the indicators (the \(f_2\left(\boldsymbol{y}_n \mid \boldsymbol{z}_n\right)\)) under the premise that different latent classes will exhibit different distributions of those indicators (\(f_1\left(\boldsymbol{y}_n \mid k\right)\)). Thus, the latent classes are designed to be “optimally different” from each other (loosely speaking) with respect to their bundle of means on the indicator vector \(yn\) – their “cluster centroids”, in the language of deterministic cluster analysis. Accordingly, the mean indicator vectors for each class offer a key basis for interpreting the class. Based on the LCCA results, we characterize the latent population segments by a variety of factors such as sociodemographic variables, education level, and AI knowledge level.

\section{Descriptive results}

This section presents descriptive results on survey responses regarding how transportation professionals perceive the efficiency and equity impacts of AI applications in transportation. As shown in Table~\ref{des1}, the survey questionnaire includes a total of 14 attitudinal statements related to AI's efficiency and equity impacts for which respondents were asked to select one of five response categories: strongly disagree, somewhat disagree, neither agree nor disagree, somewhat agree, and strongly agree. 
\begin{sidewaystable}
\scriptsize

\renewcommand\arraystretch{1.5}	
\caption{Survey responses on attitudinal statements regarding AI efficiency and equity impacts}\label{des1}
\begin{tabular}{llllll}
 \hline
 & \textbf{\makecell[l]{Strongly \\disagree}}  & \textbf{\makecell[l]{Somewhat \\disagree}} & \textbf{\makecell[l]{Neither agree \\nor disagree}} & \textbf{\makecell[l]{Somewhat \\agree}} & \textbf{\makecell[l]{Strongly \\agree}} \\
AI can lead to more efficient transportation services and cost-savings & 3.18\% & 6.01\% & 13.43\% & 53.00\% & 24.38\% \\
AI can help transportation agencies make smart, data-driven decisions & 3.87\% & 6.34\% & 11.27\% & 46.13\% & 32.39\% \\
AI can automate routine tasks and improve labor productivity  & 3.16\% & 4.91\% & 5.96\% & 43.86\% & 42.11\% \\
AI can improve traveler experience with personalized recommendations & 2.82\% & 5.63\% & 14.79\% & 48.59\% & 28.17\%\\
AI can remove bias in government decision making processes & 18.60\% & 30.88\% & 23.86\% & 20.70\% & 5.96\% \\
AI can facilitate the discovery of solutions to improve transport equity & 7.75\% & 15.14\% & 31.69\% & 35.92\% & 9.51\% \\
I believe that AI algorithms will exaggerate inequalities in\\ transportation & 5.13\% & 15.75\% & 38.83\% & 31.50\% & 8.79\% \\
Proper use of AI can help reduce social inequality & 4.74\% & 11.68\% & 35.40\% & 37.59\% & 10.58\% \\
Applying AI in transportation decision-making will reduce transparency & 8.76\% & 21.53\% & 21.90\% & 35.77\% & 12.04\% \\
Community engagement is important when developing AI\\ transportation systems & 0.73\% & 5.86\% & 11.72\% & 30.04\% & 51.65\% \\
The current AI development and deployment progress has not done\\ enough on engaging communities and disadvantaged populations & 2.19\% & 3.28\% & 30.66\% & 29.20\% & 34.67\% \\
There is limited understanding of AI ethics in the transportation\\ community & 1.09\% & 3.27\% & 8.00\% & 37.09\% & 50.55\% \\
The data used in AI applications are often not representative of the\\ population & 1.83\% & 9.16\% & 43.96\% & 26.01\% & 19.05\% \\
Biased datasets used for developing AI systems will lead to social\\ inequalities & 3.65\% & 4.01\% & 18.25\% & 33.58\% & 40.51\%\\
\hline
\end{tabular}
\end{sidewaystable}
%The next part of the descriptive analysis investigates the perception of transportation professionals as a whole. The data is derived from two sets of questions: one comprising six questions related to AI efficiency, and the other comprising eight questions related to AI equity.\textbf{ Compared to the subsequent LCCA analysis, we retain more detailed questions, allowing us to gain an overall understanding of their attitudes toward specific issues.} The results are presented in Table~\ref{des1}.

Most respondents (77\%) believe AI can enhance transportation efficiency and reduce costs. In general, a large majority of respondents (about 80\%) agree that AI can enable smart, data-driven decisions for transportation agencies and that AI can be used to automate tasks and enhance productivity. Additionally, most respondents (76\%) acknowledge AI's potential to enhance traveler experience with personalized services. However, there's limited trust that AI can eliminate bias in government decision-making processes, with only about a quarter of the respondents strongly or somewhat agreeing with this statement. 

Opinions are mixed about the potential of leveraging AI to find solutions for improving transport equity. While the number of respondents agreeing with this potential is slightly larger than those disagreeing, the two most selected response categories were ``somewhat agree" (\%36) and ``neither agree nor disagree" (32\%). The survey results also show a mixed response regarding AI's impact on transportation equity. Close to 40\% of respondents strongly or somewhat agree that AI algorithms will exaggerate inequalities in transportation, with another 39\% neither agreeing nor disagreeing with this statement. Interestingly, when asked if they agree proper use of AI can help reduce social inequality, close to half of the respondents selected ``strongly agree" or ``somewhat agree", and about 35\% selected ``neither agree nor disagree." These results seem to imply that most respondents do not believe AI carries equity benefits or harms in itself: AI's equity impacts lie in how the AI systems are developed and applied in practice. Finally, 48\% of the survey respondents agree that applying AI algorithms in decision-making will reduce transparency, 30\% of them disagree with this statement, and the remaining 22\% neither agree nor disagree. 

The last five statements are about ethical and equity considerations for AI applications in transportation. The survey results indicate overwhelming support for community engagement in AI development, with over half of respondents ``strongly agree" and another 30\% ``somewhat agree" that community engagement is important when developing AI transportation systems. However, most respondents (64\%) hold the belief that the current AI development and deployment progress has not done enough to engage communities and disadvantaged populations. Moreover, 88\% of the respondents believe that there is a limited understanding of AI ethics in the transportation community, which indicates a clear need for education and training opportunities in this regard. Concerns about data representativeness in AI applications are relatively neutral: while 45\% of the respondents agree that the data used are often not representative of the population, 44\% of them neither agree nor disagree with this statement. However, a large majority of respondents (74\%) agree that biased datasets used in AI development will lead to social inequalities.

We further conducted a series of cross-tabulations to explore the association between people's responses to the 14 attitudinal statements shown in Table \ref{des1} and their personal characteristics. We found statistically significant chi-square test results for pairs involving AI knowledge level, education level and background, and sociodemographic information (age, gender, income) with at least one of the 14 questions. Specifically, our findings indicate that younger individuals are more likely to agree that AI can improve labor productivity. Those with higher education levels, however, question AI's ability to remove bias in government decision-making processes and express reservations about the level of development in AI. As AI knowledge increases, there is a greater tendency to acknowledge that the proper use of AI can help reduce social inequality. Women and lower-income groups place more emphasis on the importance of community engagement in the development of AI transportation systems. Compared to STEM individuals, those in Non-STEM fields seem to be more sensitive to biased datasets.  Consequently, we included these characteristics as covariates of interest in the LCCA model.Note that although we did not identify statistically significant correlations between race and any of the statements, we opted to include it as one of the variables. This decision stems from the acknowledgment of the potential impact of race in some previous research.

\section{Latent Class Cluster Modeling and Results}
\subsection{Model formulation}
The first part of this section is a detailed description of the LCCA model formulation. We will introduce the structure of the two sub-models: the \textbf{membership model} and the \textbf{measurement model}. The \textbf{measurement model} utilizes latent class membership to capture associations among indicators that measure individuals' perception of AI's efficiency and equity impacts in transportation. As discussed above, the survey elicited responses to 14 attitudinal statements related to AI's impacts; since many of these statements are highly correlated, we employed factor analysis to identify a subset of indicators from them. As shown in Table~\ref{fac}, the preferred fact analysis results in a total of three factors. It is evident from the factor loadings that Factor 1 is associated with efficiency, while Factors 2 and 3 are related to equity. By considering the direction and magnitude of factor loadings and the corresponding attitudinal statement, we labeled the three factors as the following: ``AI efficiency impact", ``AI ethical concern", and ``AI equity impact." All three indicators are continuous variables. The three latent constructs were used as indicators in the LCCA measurement model.\footnote{While 359 individuals responded to the survey, only 270 of them provided complete responses for all 14 statements used to identify the three latent constructs. Therefore, the sample size for the LCCA model is N=270.}

\begin{sidewaystable}
\scriptsize
\caption{Coefficients of factors}\label{fac}
\renewcommand\arraystretch{2}	
\begin{tabularx}{\textwidth}{lccc}
\hline
 \makecell[l]{ \textbf{Question statement}}& \makecell[l]{\textbf{ Efficiency Impact}} & \makecell[l]{ \textbf{Ethical Concern}} & \makecell[l]{  \textbf{Equity Impact}}   \\\hline
\makecell[l]{AI can lead to more efficient transportation services  and cost-savings} & 0.74 &  &     \\
\makecell[l]{AI can   help transportation agencies make  smart, data-driven decisions} & 0.71 &  &    \\
\makecell[l]{AI can automate   routine tasks and improve labor productivity} & 0.79 &  &    \\
\makecell[l]{AI can remove   bias in government decision-making processes} &  &  & 0.51   \\
\makecell[l]{AI can   facilitate the discovery of solutions to improve transport equity} & 0.47 &  & 0.63   \\
\makecell[l]{AI can improve   traveler experience with personalized\\ recommendations/services} & 0.69 &  &   \\
\makecell[l]{I believe that AI algorithms will exaggerate inequalities in   transportation} &  & 0.54 & -0.41   \\
\makecell[l]{Applying AI in transportation decision-making will reduce   transparency} &  &  & -0.41   \\
\makecell[l]{Community engagement is important when developing AI   \\transportation systems} &  & 0.5 &  \vspace{0.2cm}  \\

\makecell[l]{There is limited understanding of AI ethics in the   transportation\\ community} &  & 0.49 &   \\
\makecell[l]{Proper use of AI can help reduce social inequality}  &  &  & 0.53  \\
\makecell[l]{The data used in AI applications are often not representative   of the \\population} &  & 0.49 &  \vspace{0.2cm}  \\

\makecell[l]{The current AI development and deployment progress has not done enough\\ on engaging communities, especially the disadvantaged populations} &  & 0.75 &  \vspace{0.2cm}  \\

\makecell[l]{Biased datasets used for developing AI systems will lead to \\ social inequalities} &  & 0.69 &  \vspace{0.2cm} \\\hline
\end{tabularx}
{\raggedright Note: Factor loadings below 0.3 were omitted here.}
\end{sidewaystable}

The \textbf{membership model} uses a set of covariates to predict the latent class membership, representing the latent AI perception groups in our study. As shown in Table ~\ref{var}, the covariates considered in the membership model include AI knowledge level, education level, age, gender, income, race, and educational background. Age is a binary variable where values are coded as 1 for individuals aged 40 and above, and 0 for those below 40. Race is a binary variable categorized as "Non-White" and "White". Gender is treated as a binary variable (male=1, female=0), with other options (including ``non-binary/gender non-conforming", ``not listed", or ``prefer not to disclose") treated as missing values.\footnote{The LCCA model allows for a small number of missing values. The proportions of missing values for each covariate can be found in Table~\ref{mean}.} Income, represented on a scale of 1-7, is defined as a numeric variable. We measure respondents' AI knowledge level through their familiarity with two aspects of AI: AI Concepts (machine learning, deep learning, neural networks, and reinforcement learning) and AI Technologies (encompassing computer vision, natural language processing, robotic systems, and predictive analytics). Respondents were asked to report their knowledge level of each item on a scale of 1-5 (with 1 indicating "no knowledge" and 5 indicating "expert-level knowledge"). AI knowledge level is calculated as the sum of these two self-reported values. Education background is defined as a nominal variable with four categories: "Transportation/Civil Engineering," "Urban Planning," "Other STEM majors," and "Non-STEM majors." We followed the X classification to separate the STEM (Science, Technology, Engineering, and Mathematics) and Non-STEM majors. Education Level is defined as a binary variable indicating if the respondent has a postgraduate degree (e.g., MA, MS, Ph.D., MD, JD). We did not follow the common practice of using bachelor's degree as the threshold here because 96\% of our survey respondents have a bachelor's degree.

\begin{sidewaystable}
\scriptsize
\caption{Variable description}\label{var}
\renewcommand\arraystretch{1.5}	
\begin{tabularx}{\textwidth}{ll}
\hline
\textbf{Variable}                        & \textbf{Description}                                                                                                                         \\
\hline
Factor 1: AI efficiency impact & Indicate people’s perception about AI’s impact on transportation efficiency    \\
Factor 2: AI ethical concern     & Indicate people’s concern about AI’s effects on current transportation equity    \\
Factor 3: AI equity impact     & Indicate people’s perception about AI’s impact on transportation equity  \\
\makecell[l]{AI knowledge level\\\\\\}                 & \makecell[l]{Indicate the level of knowledge in AI concepts and AI technologies . \\     1-5: “no knowledge” to “expert-level knowledge” \\ This variable is the mean of answers. So the value can be a non integer.}                               \\
\makecell[l]{Gender \\\\\\}                         & \makecell[l]{Indicate whether the person is a male or female.   \\ 1 = Male    \\ 0 = Female}                                                                                                                                                                                                          \\
\makecell[l]{Age  \\\\\\}                      & \makecell[l]{Indicate which age category the person belongs to.    \\ 0 = 18-39\\    1 = 40 and over}                                                                                                         \\
\makecell[l]{Income\\\\\\\\\\\\\\\\      }                    & \makecell[l]{Indicate which household income category the person belongs to.   \\ 1 = Less than \$25,000    \\ 2 = \$25,000–\$49,999    \\ 3 = \$50,000–\$74,999     \\ 4 = \$75,000–\$99,999     \\ 5 = \$100,000–\$124,999     \\ 6 = \$125,000–\$149,999     \\ 7 = \$150,000 or more}    \\
\makecell[l]{Race\\\\\\}                            & \makecell[l]{Indicate whether the person is White or non-White.    \\ 1 = White    \\ 0 = Non-White}                                                                                                                                                                                                    \\
\makecell[l]{Education level\\\\\\}                 & \makecell[l]{Indicate whether the person has a post-graduate degree(e.g., MA, MS,   Ph.D., MD, JD)    \\ 1 = Yes    \\ 0 = No}\\

\makecell[l]{Interaction term}                 & \makecell[l]{This variable is the product of age and AI knowledge level.    } \\                                                                                                                                 
\makecell[l]{Education background\\\\\\\\\\}                           & \makecell[l]{Indicate which field of study the person’s major belongs to.     \\ 1 = Transportation/Civil engineering   \\ 2 = Urban Planning    \\ 3 = Other STEM majors    \\ 4 = Non-STEM majors}     \\     \hline                                                                     
\end{tabularx}
\end{sidewaystable}
%Move details about the model specification and variables to a later section. Need to have detailed descriptions of several key variables (e.g., Age, Knowledge level, education background) and say something like "other variables and their values are self-explanatory."

%We hypothesized that Xs are potential covariates… After testing a variety of model specifications, we found XXX to be active covariates and YYY to be inactive.)

\begin{figure}[h!]
  \centering
  \includegraphics[width=1\textwidth]{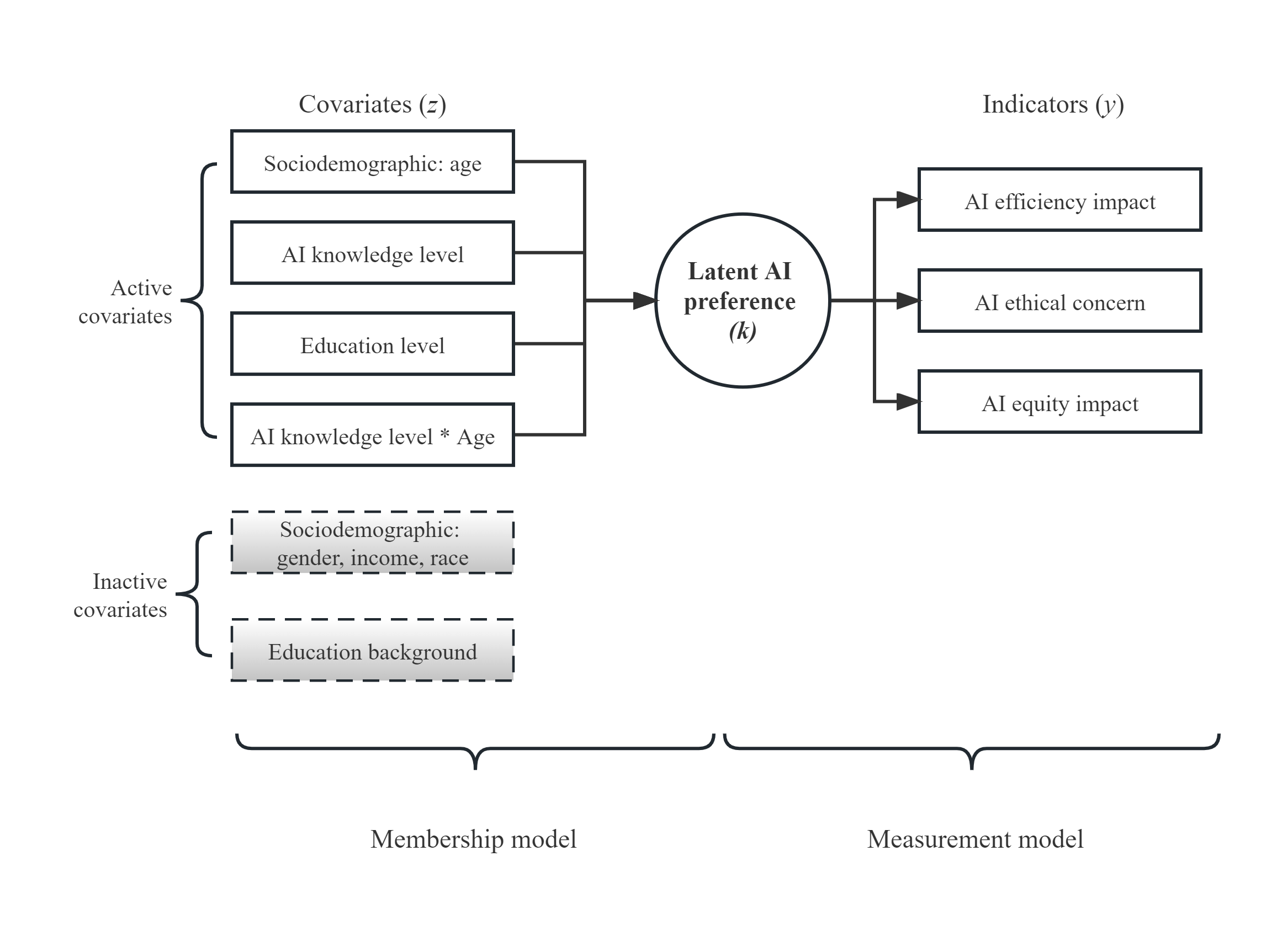}
  \caption{Model structure of the latent class cluster analysis}\label{model}
\end{figure}

After testing a variety of model specifications, we identified three active covariates for the \textbf{membership models}, including AI knowledge level, age, and education level. We further explored the presence of interaction effects and found such effects between age and AI knowledge level. We retained the insignificant covariates in the membership model as inactive covariates. The final model structure is shown in Figure~\ref{model}. 

\subsection{Identification and description of latent classes}

When deciding the final model, a major consideration involves determining the number of latent classes. This decision is usually made through a joint consideration of statistical fit and model interpretability. Models exhibiting better fit (lower Bayesian Information Criterion (BIC) values) indicate the presence of 3 or 4 latent segments in the data. However, we observed that the ``AI ethical concern" factor was statistically insignificant in the model with 3 segments, with its contribution to explaining the total variances being close to 0. In other words, the factor that measures individuals' attitudes on AI ethics and equity was not important for classifying respondents in the model comprising three latent classes. This is not ideal because we believe the inclusion of this factor in the final model adds more nuances to population segmentation and interpretation of the model results. Therefore, we opted for the model with 4 latent classes, whose model outputs are presented in Table~\ref{coef}. 

\begin{sidewaystable}
\caption{Coefficients and z-values of the estimated LCCA model (N = 270)}\label{coef}
\scriptsize
\renewcommand\arraystretch{1.5}	
\begin{tabular}{llllllllll}

\hline
\textbf{Measurement model} & \multicolumn{2}{l}{\textbf{AI Neutral}} & \multicolumn{2}{l}{\textbf{AI Optimist}} & \multicolumn{2}{l}{\textbf{AI Pessimist}} & \multicolumn{2}{l}{\textbf{AI Skeptic}} &         \\ \hline
                                      \textbf{Indicators} & Coef.          & z.             & Coef.           & z.              & Coef.          & z.             & Coef.          & z.            & p-value \\

AI efficiency impact                  & \textbf{0.3602}         & 3.9653         & \textbf{1.0786 }         & 11.5757         & \textbf{-1.2719 }       & -5.3310        & -0.1669        & -1.0665       & 1.6e-79 \\
AI ethical concern                      & \textbf{0.2758}         & 2.0103         & 0.1664          & 1.1236          & \textbf{0.5388 }        & 3.5225         & \textbf{-0.9809 }       & -2.6918       & 0.0052   \\
AI equity benefit                      & -0.0689        & -0.6070        & \textbf{0.5383 }         & 4.9827          & 0.0162        & 0.1168        & -0.4856        & -1.8579       & 6.7e-8  \\
\hline
\textbf{Intercepts}                             & Overall        & z.             & p-value         &                 &                &                &                &               &         \\
AI efficiency impact                  & \textbf{-0.2689 }       & -2.9953        & 0.0027          &                 &                &                &                &               &         \\
AI ethical concern                      & -0.1740        & -1.4021        & 0.16            &                 &                &                &                &               &         \\
AI equity benefit                      & -0.0618        & -0.6847        & 0.49            &                 &                &                &                &               &         \\
\hline
\textbf{Membership Model}  & \multicolumn{2}{l}{\textbf{AI Neutral}} & \multicolumn{2}{l}{\textbf{AI Optimist}} & \multicolumn{2}{l}{\textbf{AI Pessimist}} & \multicolumn{2}{l}{\textbf{AI Skeptic}} &  \\
\hline

\textbf{Covariates}                             & Coef.          & z.             & Coef.           & z.              & Coef.          & z.             & Coef.          & z.            & p-value \\
AI knowledge level                     & -0.2805        & -1.3161        & -0.0040          & -0.0176          & \textbf{-0.9367}        & -2.7605        & \textbf{1.2212}         & 2.5705        & 0.015   \\
Age                        &                &                &                 &                 &                &                &                &               &         \\
\makecell[r]{People under the age of 40}                                      & 0.0320        & 0.0685        & 0.7641          & 1.5803          & \textbf{1.2368}        & 2.2896        & -2.0329         & -1.8589        & 0.045   \\
\makecell[r]{People who are 40 or over}                                      & -0.0320        & -0.0685        & -0.7641          & -1.5803          & \textbf{-1.2368}        & -2.2896        & 2.0329         & 1.8589       &         \\
Age*AI knowledge level                     & -0.3187        & -1.0512        & 0.1268          & 0.4301          & \textbf{1.0392}        & 2.5937        & -0.8472         & -1.5771        & 0.037   \\
Education level                        &                &                &                 &                 &                &                &                &               &         \\
\makecell[r]{Without a postgraduate degree}                                      & \textbf{-0.4852}        & -3.1563        & 0.1990          & 1.2946          & -0.0315        & -0.1638        & 0.3177         & 1.2517        & 0.0071   \\
\makecell[r]{With a postgraduate degree}                                       & \textbf{0.4852}        & 3.1563        & -0.1990          & -1.2946          & 0.0315        & 0.1638        & -0.3177         & -1.2517           \\ \hline
\textbf{Intercept}                              & Coef.          & z.             & Coef.           & z.              & Coef.          & z.             & Coef.          & z.            & p-value \\
                                       & \textbf{2.0596}         & 4.3166         & 0.4743          & 0.9078          & 0.8045        & 1.4203        & \textbf{-3.3383}        & -2.8364       & 0.00019  \\

\hline
\end{tabular}
\end{sidewaystable}

\begin{figure}[ht]
  \centering
  \includegraphics[width=1\textwidth]{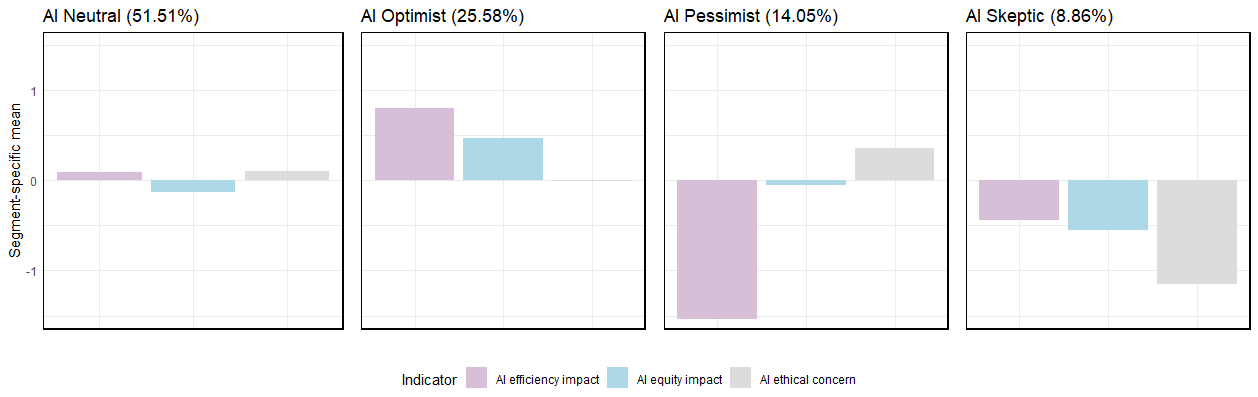}
  \caption{Mean of indicators of the four segments}\label{ident}
\end{figure}

\begin{figure}[ht]
  \centering
  \includegraphics[width=1\textwidth]{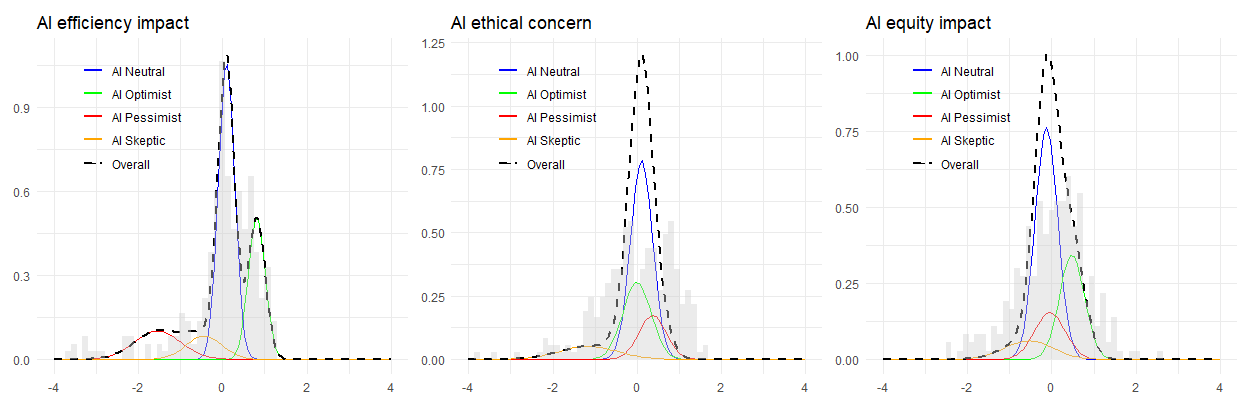}
  \caption{Sample histogram and cluster distributions of indicators (weighted N = 270)}\label{gmm}
\end{figure}
We now discuss the four latent classes based on the mean values of three indicators: \textit{AI efficiency impact}, \textit{AI ethical concern}, and \textit{AI equity impact}. Figure~\ref{ident} presents the segment-specific mean values of these indicators. It is important to note that the sample means of the indicators are all close to zero, as they are generated through standardized factor analysis. A value of zero for each factor indicates a neutral perspective. Thus, the sign and magnitude of the segment-specific mean can be directly utilized to distinguish attitudes. For \textit{AI efficiency impact} and \textit{AI equity impact}, positive values would indicate benefits whereas negative values would indicate no benefits or even harms. In other words, a negative value associated with these factors would indicate that the respondent believes AI applications in transportation have few to no efficiency or equity benefits. For \textit{AI ethical concerns}, larger values indicate greater levels of concern for AI's ethical and equity implications, whereas negative values can be interpreted as the respondent having little of such concerns.

%should be understood as "The current AI development and deployment progress has not done enough on readiness in terms of AI equity", whereas negative values should be understood as a potential indication of greater satisfaction or confidence in AI equity readiness.} 

%It is worth emphasizing that a negative value should be interpreted as indicating that the segment perceives the absence of a corresponding effect (e.g., perceiving no efficiency benefits from AI), rather than suggesting that the segment believes AI has opposing effects. 

\textbf{Segment 1} constitutes the largest portion of the survey respondents (51.38\%). For this group, the mean values of all three factors are close to 0, which means that they tend to have a neutral perception of AI's efficiency and equity impacts. Therefore, we label this segment as "\textbf{AI Neutral}". %or AI neutral? (from tom)

\textbf{Segment 2} accounts for 25.48\% of the survey sample. In this segment, factors \textit{AI efficiency impact} and \textit{AI equity impact} display significantly positive values, which indicates that this group believes that AI applications in transportation can deliver both efficiency and equity benefits. The mean value of factor \textit{AI ethical concern} is close to zero, indicating a moderate level of concern for the ethical and equity issues associated with AI. We label this segment as "\textbf{AI Optimist}".%or AI optimists?(from tom)

%It is worth noting that not only the mean value of AI ethical concern in \textbf{AI Neutral} is approximately zero, but \textbf{AI Optimist} also demonstrates an almost negligible mean value (\textbf{though a negative value is more intuitive for the "enthusiast" label}). These observations suggest that \textbf{AI Optimist} maintain neutral stances concerning the readiness related to AI equity. \textbf{It can be further inferred that this aspect has yet to receive full consideration from them at present.}

\textbf{Segment 3} represents 13.68\% of the respondents. Within this segment, the mean value of factor \textit{AI efficiency impact} is negative and the mean value of factor \textit{AI equity impact} is close to zero. This suggests that this group believes that AI applications in transportation will not bring efficiency benefits, and they hold a neutral stance regarding AI's equity impacts. Moreover, the mean value of factor \textit{AI ethical concern} is positive, which means that respondents in this segment have significant concerns about the ethical and equity implications of AI use in transportation. Based on these interpretations, we refer to this segment as "\textbf{AI Pessimist}". %are they necessarily opposed to AI or just express unfavorable opinions?(from tom)

\textbf{Segment 4} encompasses 9.46\% of the population, and all average values within this group are negative. This group has serious doubts about the benefits of AI, and they have little concern that AI use in transportation will cause major ethical and equity issues. We interpret these results as suggesting that this group does not believe that AI will make a major impact on the field of transportation. We designate this segment as "\textbf{AI Skeptic}".

Figure~\ref{gmm} illustrates the composite distributions of the three indicators alongside the segment-specific distributions. The sample histograms of the indicators are depicted as gray bars. The black dashed lines represent the 'overall' distributions, which are probability-weighted superpositions of the four constituent normal distributions. This visualization effectively separates and elucidates the distinct contributions of each latent segment, providing insights into the highly irregular distributions of mode-use propensities. The LCCA results demonstrate a notable approximation to the sample distributions, further highlighting the efficacy of the latent class analysis in capturing the underlying patterns.

\subsection{Segment profiles}
In this section, we present segment-specific profiles for the four identified segments. The distributions of active and inactive covariates are summarized in Table~\ref{mean}. Note that the sampled respondents had a high level of education overall, with 71\% of respondents holding post-graduate degrees. Throughout the analysis, the education level is measured based on the proportion of each segment holding post-graduate degrees. Therefore, when stating ``lower education levels," we refer to a comparison with the sample mean rather than with the general public. Additionally, it should be noted that the AI knowledge level discussed here is self-assessed knowledge about AI concepts and technologies.

\begin{sidewaystable}
\caption{Segment-specific means/shares of indicators and covariates (N = 270)}\label{mean}
\scriptsize
\renewcommand\arraystretch{1.1}	

\begin{tabular}{llllll}
\hline
                                           & \textbf{AI Neutral} & \textbf{AI Optimist} & \textbf{AI Pessimist} & \textbf{AI Skeptic} & \textbf{Overall} \\
\hline
\textbf{Cluster Size}                      & 0.5151               & 0.2558                 & 0.1405            & 0.0886              &                  \\
\hline
\textbf{Indicators}                        &                      &                        &                      &                     &                  \\
\textbf{AI efficiency impact}             & 0.0912               & 0.8097                 & -1.5408              & -0.4359             & -0.0011          \\
\textbf{AI ethical concern}                 & 0.1018               & -0.0076                & 0.3648               & -1.1549             & -0.0005          \\
\textbf{AI equity impact}                 & -0.1307              & 0.4765                 & -0.0455              & -0.5474             & -0.0003          \\
\hline
\textbf{Active covariates}                 &                      &                        &                      &                     &                  \\
\textbf{AI knowledge level}                &                      &                        &                      &                     &                  \\
\makecell[r]{2}                                          & 0.2186               & 0.0782                 & 0.2186               & 0.1630              & 0.1778           \\
\makecell[r]{3}                                    & 0.1104               & 0.1150                 & 0.1224               & 0.0438              & 0.1074           \\
\makecell[r]{4 - 5}                                        & 0.3469               & 0.4242                 & 0.2688               & 0.1765              & 0.3406           \\
\makecell[r]{6}                                          & 0.1920               & 0.1599                 & 0.2352               & 0.2652              & 0.1964           \\
\makecell[r]{7 - 10}                                    & 0.1321               & 0.2226                 & 0.1550               & 0.3515              & 0.1779           \\
\makecell[r]{Mean}                                       & 2.2206               & 2.5669                 & 2.2705               & 2.9402              & 2.3799           \\
\textbf{Education level}                   &                      &                        &                      &                     &                  \\
\makecell[r]{Without a postgraduate degree}                                          & 0.1912               & 0.3929                 & 0.3351               & 0.4428              & 0.2853           \\
\makecell[r]{With a postgraduate degree}                                          & 0.8018               & 0.6070                 & 0.6646               & 0.5567              & 0.7110           \\
\makecell[r]{Prefer not to answer}        & 0.0070               & 0.0000                 & 0.0003               & 0.0005              & 0.0037           \\
\textbf{Age}                               &                      &                        &                      &                     &                  \\
\makecell[r]{People under the age of 40}                                      & 0.4593               & 0.5350               & 0.3143               & 0.0912              & 0.4257           \\
\makecell[r]{People who are 40 or older}                                         & 0.4788               & 0.4097                 & 0.6381               & 0.7932              & 0.5113  
\\
\makecell[r]{Prefer not to answer}                                         & 0.0620               & 0.0553                 & 0.0476               & 0.1157              & 0.0630    \\
\textbf{Age*AI knowledge level}                               &                      &                        &                      &                     &                  \\
\makecell[r]{Mean}                                      & 1.0482               & 1.0777               & 1.6722               & 2.2718              & 1.2518           \\
\hline
\textbf{Inactive covariates}                        &                      &                        &                      &                     &                  \\
\textbf{Gender}                            &                      &                        &                      &                     &                  \\
\makecell[r]{Female}                                          & 0.2992               & 0.3136                 & 0.2685               & 0.1948              & 0.2888           \\
\makecell[r]{Male}                                          & 0.6388               & 0.6585                 & 0.6311               & 0.6567              & 0.6444           \\
\makecell[r]{Other} & 0.0620               & 0.0279                 & 0.1004               & 0.1485              & 0.0667           \\
\textbf{Income}                            &                      &                        &                      &                     &                  \\
\makecell[r]{Less than \$74999}                                      & 0.1487               & 0.1455                 & 0.1646               & 0.1677              & 0.1519           \\
\makecell[r]{\$75000-\$99999}                                          & 0.1472               & 0.1361                 & 0.1208               & 0.1466              & 0.1407           \\
\makecell[r]{\$100000-\$124999}                                          & 0.1505               & 0.1247                 & 0.2874               & 0.0755              & 0.1556           \\
\makecell[r]{\$125000-\$149999}                                          & 0.1366               & 0.1528                 & 0.0726               & 0.1507              & 0.1333           \\
\makecell[r]{\$150000 or more}                                          & 0.2579               & 0.3155                 & 0.2587               & 0.2330              & 0.2704           \\
\makecell[r]{Prefer not to answer}        & 0.1589               & 0.1254                 & 0.0959               & 0.2265              & 0.1482           \\
\makecell[r]{Mean}                                       & 5.1003               & 5.2914                 & 5.0097               & 5.1227              & 5.1387           \\
\textbf{Race}                              &                      &                        &                      &                     &                  \\
\makecell[r]{Non-white}                                          & 0.2906               & 0.3696                 & 0.2863               & 0.1434              & 0.2962           \\
\makecell[r]{White}                                          & 0.6384               & 0.5652                 & 0.6118               & 0.5852              & 0.6111           \\
\makecell[r]{Prefer not to answer}        & 0.0710               & 0.0652                 & 0.1020               & 0.2714              & 0.0927           \\
\textbf{Educational background}              &                      &                        &                      &                     &                  \\
\makecell[r]{Transportation/Civil engineering}                                          & 0.4555               & 0.5425                 & 0.3964               & 0.4243              & 0.4666           \\
\makecell[r]{Urban planning}                                          & 0.2615               & 0.1991                 & 0.2391               & 0.0453              & 0.2221           \\
\makecell[r]{Other STEM majors}                                          & 0.1161               & 0.1378                 & 0.1772               & 0.2309              & 0.1408           \\
\makecell[r]{Non-STEM majors}                                          & 0.1256               & 0.0868                 & 0.1042               & 0.1472              & 0.1148           \\
\makecell[r]{Prefer not to answer}        & 0.0413               & 0.0338                 & 0.0831               & 0.1523              & 0.0556         \\ 
\hline
\end{tabular}
\end{sidewaystable}

The most prominent characteristic of \textit{AI Neutral} is their exceptionally high level of education compared to the other three segments. Within this segment, 80\% of individuals possess post-graduate degrees, making them the group with the highest level of education. Simultaneously, this segment exhibits the lowest AI knowledge level, with 68\% of respondents indicating a self-reported AI knowledge level below 3 (where 3 signifies a moderate level of knowledge). Moreover, \textit{AI Neutral} is the second youngest among the four segments, being only slightly older than \textit{AI Optimist}. In summary, \textit{AI Neutral} comprises younger individuals with very high levels of education and lower levels of AI knowledge.

The most prominent characteristic of \textit{AI Optimist} is its notably young age, with 54\% of these individuals falling under the age of 40. Moreover, the education level of \textit{AI Optimist} ranks second lowest among the four segments, with 61\% holding post-graduate degrees, which is significantly lower than the overall sample mean of 71\%. However, their level of knowledge about AI slightly surpasses the sample average. In addition, the proportion of white people in this segment is the lowest (57\%). 54\% of individuals in this segment major in transportation/civil engineering, which is the highest among the four segments. In summary, \textit{AI Optimist} comprises the youngest individuals with relatively lower levels of education and relatively higher levels of AI knowledge.

The \textit{AI Pessimist} exhibits the second-highest average age among the four segments. While their education level and AI knowledge level are both slightly lower than the sample mean, the discrepancies are marginal. In contrast to the other three segments, the predominant income range within this segment falls between \$100,000 and \$124,999, rather than \$150,000 or more. In summary, \textit{AI Pessimist} encompasses older individuals with moderate levels of education and AI knowledge.

The \textit{AI Skeptic} has the lowest level of education, with only 56\% holding post-graduate degrees, which is 25\% lower than \textit{AI Neutral}. Interestingly, this segment displays the highest AI knowledge level. Merely 38\% of \textit{AI Skeptics} believe that their AI knowledge level is below 3, which is 29\% percentage points lower than that of \textit{AI Neutrals}. Additionally, the average age of \textit{AI Skeptics} is the highest among the four segments. Only 9\% of people in this segment are below 40 years old, while in the entire sample, this number is 43\%. 5\% of individuals in this segment major in urban planning, which is the lowest among the four segments. Overall, \textit{AI Skeptics} encompass older individuals with lower levels of education, yet with higher levels of AI knowledge.

\subsection{Impacts of active covariates on latent class membership}

In this section, we discuss the distribution of latent class membership conditioning on the values of active covariates: age, education level, and AI knowledge level.  

%the impact of covariates. The primary approach is "pairwise comparisons." Specifically, we will compare the means and distributions of active covariates within segments to further delve into the intricate relationship between the covariate values and segment characteristics. Through this analyzing process, we can also enhance our understanding of the characteristics of segments. 

%We have previously mentioned the necessity of identifying an interaction term based on Age and AI knowledge level. However, comparing the means of this interaction term is challenging to interpret. Therefore, we have employed an auxiliary method: analyzing membership models. Here, we examine the coefficients derived and explore the results of statistical significance. During this process, we can observe how active covariates influence the probability of a respondent being classified into a specific segment. By combining these two methods, we can ensure both model exploration and the validity of our findings.

%\begin{figure}[ht]
  %\centering
  %\includegraphics[width=0.57\textwidth]{age.png}
  %\caption{Age}\label{age}
%\end{figure}

Previously we saw that the \textit{AI Neutral} and \textit{AI Optimist} segments tend to be younger compared to the \textit{AI Pessimist} and \textit{AI Skeptic} segments. In fact, we found that 88\% of the respondents who are under age 40 belong to the \textit{AI Neutral} or \textit{AI Optimist} segments whereas less than 2\% of them are in the \textit{AI Skeptic} segment. Moreover, the percentage of respondents who are 40 or older classified as an \textit{AI Skeptic} is nearly double of the percentage classified as an \textit{AI Optimist}. These results suggest that, compared to older professionals, younger individuals tend to hold more optimistic or neutral attitudes toward AI use in transportation and are much less likely to be an \textit{AI Skeptic}. This finding is consistent with common perception: younger people tend to have more positive views of new technologies and expect more significant impacts from them.

On education level, we observed that the percentage of respondents holding a postgraduate degree is much higher in the \textit{AI Neutral} segment than in other segments. Outputs of the membership model confirm that a higher education level is significantly associated with a greater probability of being an \textit{AI Neutral}. Our interpretation is that people with postgraduate education tend to make more cautious judgments regarding the potential impacts of AI use in transportation. 

%So compared to the other three segments holding a more progressive attitude towards AI,  the \textbf{AI Neutral} chooses to adopt a neutral and conservative stance in this stage of AI development.

%\begin{figure}[ht]
  %\centering
  %\includegraphics[width=0.57\textwidth]{knowledge1.png}
  %\includegraphics[width=0.57\textwidth]{knowledge2.png}
  %\caption{AI knowledge level}\label{know}
%\end{figure}

Finally, regarding the impact of AI knowledge level on the probability of a respondent being classified into a specific segment, we have previously noted the importance of including an interaction term between the age variable and AI knowledge level in the membership model. This is because, through some exploratory analyses, we found the effects of AI knowledge level on latent class membership to vary across age groups. Specifically, for people under the age of 40, having a higher level of knowledge reduces the probability of them being an \textit{AI Pessimist}. For people who are 40 years old or older, AI knowledge level has nearly no effect on the likelihood of being an AI Pessimist. Moreover, the membership model shows that having a higher AI knowledge level is associated with a higher probability of being an \textit{AI Skeptic} for all age groups.

\section{Discussion}

%offer your interpretation of the finding or discuss the implications of the finding.

%While some AI transportation applications have already been implemented in the real world, many applications are still in the development or experimental stages. 

In this section, we discuss the policy and practical implications of the study results on the deployment of AI applications across the transportation sector. To better engage with the existing literature and broader conversation about AI, our discussion below may occasionally extrapolate the survey findings from transportation professionals to the general public. Specifically, we believe that results regarding the identification of latent classes and the sociodemographic profiles of each segment are most likely extrapolatable.

%whereas the key difference lies in the relative share of populations belonging to each segment. For example, we expect that share of the general public %While we expect some differences between professionals working in the transportation industry and the general public regarding their perceptions of AI, we believe that .

First of all, it appears that how people perceive AI differs from how they perceive autonomous vehicles. Even though autonomous vehicles are a major domain for AI applications, we observe some differences. Notably, while studies often identify gender differences in attitudes towards autonomous vehicles \citep{horowitz2021influences, kassens2020willingness}, we find that gender is not a significant factor in determining latent class membership. However, consistent with the existing literature that suggests a strong link between people's age and their acceptance of new technology \citep{hohenberger2016and}, we find that respondents' perception of AI's impacts on transportation correlates with their age. Our survey results suggest that, compared to people aged 40 or above, younger professionals are much more likely to be an \textit{AI Neutral} or \textit{AI Optimist} than an \textit{AI Pessimist} or \textit{AI Skeptic}. These results suggest that older adults should be the primary target for education and outreach efforts if AI applications in transportation continue to grow in significance and will likely impact all aspects of transportation.

%we find that the active covariates we identified differ from some existing research on autonomous vehicles. For instance,  However, in our study, gender was not a significant covariate. This is because we measured respondents' overall attitudes towards AI, rather than attitudes toward a specific application. Therefore, a proper way to utilize the findings of this study is to derive suggestions from the significant covariates we identified to foster a positive and inclusive environment for AI in transportation.

%(mainly for people aged 40 or above because the number of younger respondents in the \textit{AI Skeptic} segment is very small). 

Moreover, our study reveals an intricate relationship between transportation professionals' AI knowledge level and their latent class membership. Surprisingly, we have observed that \textit{AI Skeptics} have the highest AI knowledge levels among the four segments. In other words, transportation professionals who know AI better also tend to be people who doubt that AI will lead to significant impacts or that AI use in transportation will lead to major ethical concerns. A possible explanation is many respondents in the \textit{AI Skeptic} segment have encountered many AI-powered systems or technology products that brand themselves as ``AI" but have yet to see major use cases from them that lead to observable impacts; these experiences make many of them consider AI as another short-lived technology that would not make a big impact in transportation. If this explanation is proven, it means that more successful AI use cases in transportation can reshape the perceptions of \textit{AI Skeptics} over time. Moreover, we find that, for younger professionals, having a higher AI knowledge level reduces the probability of them being an \textit{AI Pessimist}. This could reflect a reverse causality problem, that is, young people who are not pessimist about AI are more likely to acquire more knowledge about it; however, this finding also implies that more AI education and training can help students and younger people better adapt to a new era increasingly shared by AI.

Finally, to promote the greater applications of AI in proper use cases to maximize their benefits, continued education and training programs are required to better prepare the transportation workforce. While transportation professionals surveyed here have a high education level overall, their self-reported AI knowledge level remains relatively low. On the one hand, this may lead to erroneous decision-making regarding AI applications. On the other hand, given the rapid development of artificial intelligence recently, individuals with a low AI knowledge level may miss out on its benefits and even risk being displaced from their jobs. As a result, low AI knowledge levels among transportation professionals could increase both societal costs and personal risks. To address this challenge, targeted AI education should be prioritized. We have observed that transportation professionals under 40 are highly unlikely to be an \textit{AI Skeptic}, but a low AI knowledge level could potentially make them an \textit{AI Pessimist}. Thus, the focus should be on enhancing incomplete or insufficient AI training and bridging the gap between AI theory and practice. For older professionals, their perception of AI may be outdated or stereotyped. Hence, updating or augmenting their knowledge should be a top priority. Professional workshops or conference sessions that demonstrate successful AI use cases would be very beneficial.

%Specifically, \textit{AI Pessimists} require familiarization of the concepts and technologies of AI, which helps them gain awareness of the potential benefits of AI. However, even with the same negative attitude, \textit{AI Skeptics} need help in building confidence in the implementation of AI. Furthermore, compared to formal AI education in schools, employers or departments may lack the incentive to provide AI education. Public funding initiatives are necessary to drive AI education in such settings.

\section{Conclusion}

This study aims to assess how transportation professionals in North America perceive AI applications in transportation and their efficiency and equity impacts. Toward this goal, we conducted a survey to ask transportation professionals about their views on AI, its potential impacts, and related ethical and equity concerns for AI use in transportation. We conducted a descriptive analysis of the survey results and applied latent class cluster analysis to identify latent population segments with distinctive perceptions toward AI's efficiency and equity impacts. 

The survey indicates widespread optimism regarding AI's potential to enhance transportation efficiency and reduce costs. Over three-quarters of the respondents believe in AI's ability to facilitate smart, data-driven decisions, automate tasks for transportation agencies, and improve traveler experiences. However, only a small minority (25\%) believe that AI can eliminate bias in government decisions. Moreover, opinions are mixed on the potential of leveraging AI to find solutions for improving transport equity, and many respondents expressed concerns about the use of AI algorithms in transportation to exaggerate existing inequalities. About half of the respondents also have concerns about the use of AI to reduce transparency in transportation decision-making. In addition, transportation professionals express strong support for community engagement in AI development, yet 64\% of them feel current efforts neglect disadvantaged communities. A large majority (88\%) perceive a limited understanding of AI ethics in the transportation sector. Finally, most respondents agree that biased datasets used to support AI development can contribute to social inequalities.

%The objective of our survey was to assess the level of segmentation among transportation professionals regarding their views on AI, its potential impacts, adoption barriers, and related ethical issues. It also evaluated respondents' AI knowledge and training, and probes the ethical considerations of AI's application within transportation. Its goal is to discern the current AI status within the sector, the obstacles to its use, the integration of equity and ethics, and the workforce's readiness for widespread AI use. 

The LCCA further helped us identify four distinctive segments and their respective shares in the survey sample, which we labeled as \textit{AI Neutral} (51.5\%), \textit{AI Optimist} (25.6\%), \textit{AI Pessimist} (14.1\%), and \textit{AI Skeptic} (8.9\%). However, one should interpret the shares of the four segments with caution because our survey sample is not representative of the whole transportation profession (highly educated individuals are overrepresented here). A further analysis of the LCCA outputs reveals the following: (1) Age significantly impacts transportation professional's potential attitudes toward AI transportation applications. Older respondents (aged 40 or above) are more likely to have negative attitudes than younger respondents (aged under 40). (2) Individuals with a post-graduate degree are more likely to hold neutral attitudes toward AI. (3) Having a higher AI knowledge level increases the possibility of a transportation professional being an \textit{AI Skeptic} and decreases the possibility for a younger professional being an \textit{AI Pessimist}.

Understanding the characteristics of particular segments of survey respondents is important when explaining survey results because it allows for a more nuanced and comprehensive interpretation of patterns of responses. Sociodemographic characteristics such as age and education level as well as levels of knowledge about AI can expose distinct trends and patterns, reveal inherent biases, and enable personalized communication strategies. Moreover, this segmented understanding of the data can guide effective policy and strategy formulation, particularly if some groups of professionals exhibit resistance or uncertainty towards AI that could be alleviated with targeted educational efforts. Additionally, these trends offer insight into potential future changes in the overall attitudes toward AI within the profession as the demographic makeup shifts over time. 

Building on our findings, future research should aim to better understand the underlying factors that influence respondent attitudes beyond demographics and AI knowledge. Qualitative methods like interviews or focus groups can be employed to gather further insights into the reasoning behind the skepticism or neutrality towards AI in transportation. Additionally, studying the impact of different forms of AI education and awareness programs on altering these attitudes can be crucial. Comparative analyses involving transportation professionals from diverse cultural and geographical backgrounds may also unveil regional variations in attitudes and AI acceptance levels. Furthermore, the role of organizational culture and policies in shaping professionals’ perspectives towards AI should be investigated. As the transportation sector continues to evolve, it is imperative to examine how organizations can foster an environment that is receptive to innovation while addressing the concerns of their employees. Experimental design can be used to assess the impact of interventions like workshops, seminars, and training sessions on AI literacy and acceptance. Ultimately, a multidimensional approach, considering psychological, organizational, and cultural aspects, will be essential to fully comprehend the intricate landscape of AI acceptance and use in the transportation field and inform the development of inclusive, effective strategies for AI integration.

\section{Acknowledgments}
The authors would like to thank Mike Hunter for providing helpful suggestions on designing the survey. We are grateful for the funding support from the Center for Equitable Transit-Oriented Communities Tier-1 University Transportation Center (Grant No. 69A3552348337) and the Southeastern Transportation Research, Innovation, Development, and Education (STRIDE) Region 4 University Transportation Center (Grant No. 69A3551747104). ChatGPT was used to improve the readability and language of the work. 

\newpage

\bibliography{sn-bibliography}% common bib file
%% if required, the content of .bbl file can be included here once bbl is generated
%%\input sn-article.bbl

\end{document}